\documentclass[conference]{IEEEtran}
\usepackage{subfigure}
\usepackage{pifont}
\usepackage{bbding}
\usepackage{times}
\usepackage{cite}
\usepackage{times}
\usepackage{amsmath}
\usepackage{amssymb}
\usepackage[dvips]{graphicx}
\usepackage{subfigure}
\usepackage{bm}
\usepackage{booktabs}
\usepackage{algorithmic}
\usepackage[lined,algonl,ruled]{algorithm2e}
\usepackage{multirow}
\makeatletter
\def\ps@headings{%
\def\@oddhead{\mbox{}\scriptsize\rightmark \hfil \thepage}%
\def\@evenhead{\scriptsize\thepage \hfil \leftmark\mbox{}}%
\def\@oddfoot{}%
\def\@evenfoot{}}
\makeatother

\ifCLASSINFOpdf
\else
\fi
\begin{document}
%
\title{YouSense: Mitigating Entropy Selfishness in Distributed Collaborative Spectrum Sensing}

\author{Shuai Li$^\dag$$^\ddag$, Haojin Zhu$^\dag$, Zhaoyu Gao$^\dag$, Xinping Guan$^\dag$, and Kai Xing$^\S$\\
$^\dag$ Shanghai Jiao Tong University, Shanghai, China\\
\{zhu-hj, zy-gao, xpguan\}@sjtu.edu.cn\\
$^\ddag$ University of Minnesota, Twin Cities, USA\\
lixx2381@umn.edu\\
$^\S$ University of Science and Technology of China, P.R. China\\
kxing@ustc.edu.cn\\
\vspace{-1cm}
\authorblockA{~~}
 }


%


\maketitle

\begin{abstract}

Collaborative spectrum sensing has been recognized as a promising approach to improve the sensing performance via exploiting the spatial diversity of the secondary users.
In this study, a new selfishness issue is identified, that selfish users sense no spectrum in collaborative sensing. For easier presentation, it's denoted as entropy selfishness. This selfish behavior is difficult to distinguish, making existing detection based incentive schemes fail to work.
To thwart entropy selfishness in distributed collaborative sensing, we propose YouSense, a One-Time Pad (OTP) based incentive design that could naturally isolate entropy selfish users from the honest users without selfish node detection. The basic idea of YouSense is to construct a trapdoor one-time pad for each sensing report by combining the original report and a random key. Such a one-time pad based encryption could prevent entropy selfish users from accessing the original sensing report while enabling the honest users to recover the report. Different from traditional cryptography based OTP which requires the key delivery, YouSense allows an honest user to recover the pad (or key) by exploiting a unique characteristic of collaborative sensing that different secondary users share some common observations on the same radio spectrum. We further extend YouSense to improve the recovery successful rate by reducing the cardinality of set of the possible pads. By extensive USRP based experiments, we show that YouSense can successfully thwart entropy selfishness with low system overhead.

\end{abstract}

{\bf \it Keywords} -- \textbf{\small Cognitive Radio Security, Collaborative Spectrum Sensing, Incentive Mechanism Design \\}

\vspace{-0.2cm}
\section{Introduction}\label{introduction}

The ever increasing spectrum demand with the emerging wireless applications has inspired the concept of Cognitive Radio (CR)\cite{begincr}, which is proposed to optimize the utilization of the precious natural resource, the radio spectrum. Different from the conventional spectrum management paradigm in which most of the spectrum is allocated to fixed licensed users or the primary users for exclusive use, a CR system allows secondary users to utilize the idle spectrum\cite{survey1}, as long as intolerable interference to primary users is not introduced.

One major challenge to the CR system is how to enable the secondary users to accurately detect the presence of the primary user. It is recently discovered that collaboration among secondary users can significantly improve the performance of spectrum sensing by exploiting their spatial diversity\cite{cosensing1}. Thus, collaborative sensing has been widely adopted in existing standards or proposals, i.e., IEEE 802.22 WRAN, CogNeA, IEEE 802.11af and WhiteFi\cite{NDSS11}. Though the FCC's recent ruling eliminates spectrum sensing as a requirement for devices that have geo-location capabilities and can access a new TV band (geo-location) database, spectrum sensing and its variants are still expected to play an important role in improving the CR network performance, since geo-location database may unavailable or inaccurate for some cases. Besides, for other bands (like the band allocated to the microphone, etc.), maintaining a database is practically infeasible\cite{kangshin}. In these cases, spectrum sensing is indispensable.


Most of collaborative sensing solutions assume that all CR users are willing to share their sensing results. This assumption, however, might be easily violated in distributed collaborative sensing for lack of a centralized authority, and selfish users refusing to multicast sensing reports may prevail. This selfish behavior reduces the cooperation diversity, and thus inevitably degrades the collaborative sensing accuracy. Existing solutions to this problem include the traditional reputation/credit based incentive mechanisms\cite{reputation1,reputation2,credit2,credit5} that are designed for traditional wireless networks, as well as the proposed CR specific incentive schemes\cite{beibei,qianzhang}.

\emph{But motivating CR users to share is one thing, getting them to sense, another.}
In this paper, a new selfishness model is identified, formulating the misbehavior of sensing no spectrum in collaborative sensing. Noticing the selfish user of this kind could share fabricated non-innovative results, existing incentive mechanisms fail to work. What makes things worse is that those fabricated results could seem normal: since CR users are simultaneously the sensing report publishers and recipients, the selfish user can impersonate an honest one by simply claiming duplicated (or slightly modified) reports from others as his own. Comparing with traditional selfishness, this misbehavior is covert and cannot be easily detected. In this paper, it's simply denoted as \emph{entropy selfishness}. Entropy selfishness is attractive to the selfish users, because on one hand it enables the selfish users to save the sensing overheads (e.g., power consumption, sensing time) or even the hardware (e.g., scanner) cost. On the other hand, detecting such entropy selfishness is much more difficult than traditional selfish behaviors (e.g., packet dropping), which makes it safe to adopt.

Entropy selfishness poses a serious threat to collaborative sensing. Since entropy selfish users make no spectrum sensing, this misbehavior may seriously reduce the sensing diversity and consequently degrade the collaborative sensing performance. This problem will be more challenging in some circumstances where spectrum sensing performance has been optimized according to the number of the collaborators. In such a case, entropy selfishness may result in undesirable consequences, such as the low primary user detection probability.\footnote{Generally speaking, the more CR users joining the collaboration, the more aggressive sensing strategies will be adopted to detect the existence of the primary user. If there exists entropy selfish users, the chosen sensing strategy may turn out to be too aggressive to detect the presence of the primary user.}

\begin{figure}
\centering
\includegraphics[width=0.4\textwidth]{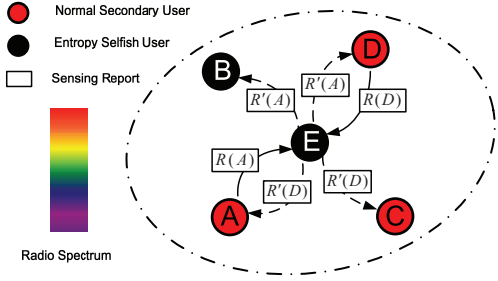}
\vspace{-0.2cm}
\caption{\textbf{Entropy Selfishness:} in collaborative spectrum sensing, the entropy selfish users doesn't make their own sensing and just forward the duplicated or slightly modified sensing reports from other secondary users. For example, the entropy selfish user $E$ copies user $D$'s report $R(D)$ and $A$'s report $R(A)$, then sends $R'(D)$ to $A$ and $C$, $R'(A)$ to $B$ and $D$, respectively.}
\vspace{-0.5cm}
\end{figure}

Unfortunately, existing incentive mechanisms fail to thwart entropy selfishness due to following technical challenges:
\textbf{Unobservability.} Different from traditional selfishness (e.g. packet dropping) which can be easily observed, detecting entropy selfishness may represent a great challenge. Because entropy selfish users do share reports, distinguishing the copied reports from the innovative ones is technically hard.\\
\textbf{Stringent Sensing Delay.} The CR networks are often characterized as delay sensitive systems, and the sensing delay is on the order of milliseconds\cite{tradeoff1}. This means the mechanism design for thwarting entropy selfishness shouldn't involve the computational expensive cryptographic operations.\\
\textbf{Distributed Fashion.} Existing incentive schemes normally assume a centralized trusted third party taking trust management or credit clearance. But, in distributed sensing such a centralized third party may not be available, and our incentive scheme should be designed in distributed fashion.

In this paper, we introduce YouSense, a One-Time Pad (OTP) based incentive design that stimulates the CR users to make their own sensing in cooperation. YouSense is not based on selfish node detection. Instead, it can prevent entropy selfish users from harvesting collaboration even without recognizing which one is of entropy selfishness. The basic idea of YouSense is to construct a trapdoor OTP for each sensing report by combining the original report with a random key (or pad). Such an OTP based encryption can prevent the entropy selfish users from accessing the original sensing results while still allowing the honest ones to recover the report. And yet, different from traditional OTP, the honest user in YouSense can recover the sensing results without key delivery. This property is due to a unique characteristic of collaborative sensing, that different users share some common observations over the same radio spectrum. We further improve the OTP method to guarantee successful recovery rate, meanwhile still keeping each bit of the OTP ciphertext leak no spectrum information to entropy selfish users. By USRP based experiment, we show YouSense can successfully thwart entropy selfishness with quite low system overhead.

The contributions of this work are summarized as follows:
\begin{itemize}
  \item We identify a new selfishness in distributed collaborative sensing, coined as entropy selfishness. Compared with the traditional selfish issues in wireless networks, the entropy selfishness is more difficult to detect and thus can't be addressed by existing incentive mechanisms.
   \item We propose YouSense to mitigate entropy selfishness in collaborative sensing. With YouSense, only honest secondary users who scan the whole spectrum can obtain the sensing results of others, while selfish secondary users can't derive the spectrum information.
  \item We also evaluate the performance of YouSense through both the theoretical analysis and USRP-based experiments. The impact of YouSense on the collaborative spectrum sensing is also investigated in this paper.
\end{itemize}

This paper is organized as follows. In section II, we give the preliminary. In section III, the basic and advanced YouSense are described. In section IV, the performance evaluation of YouSense is introduced. Section V gives the related works, and in section VI, we conclude this paper.

\section{Preliminary}\label{system model}
In this section we will describe the channel model, collaborative sensing model, and the Entropy Selfishness model.

\subsection{Channel Model}
In CR networks, a channel/spectrum could be accessed by the secondary user if it is not occupied by the primary user. Suppose there are $M$ channels $\widetilde{C} = \left\{ {\widetilde{C}_1 ,\widetilde{C}_2 , \cdots ,\widetilde{C}_M } \right\}$, and these channels are separately owned by $M$ primary users.

We model each channel as an ON-OFF source, in which ON-state indicates the channel is occupied, and OFF-state represents this channel is idle.
For ease of the presentation, let $\widetilde{C}_i=1$ denote the ON-state, and $\widetilde{C}_i=0$ denote the OFF-state\cite{hanzhu}.
Besides, we assume ON and OFF periods of the channel follow the exponential distribution with rate $\lambda_{ON}$ and $\lambda_{OFF}$ respectively\cite{Poisson2}. Without loss of the generality, we also assume $M$ channels in this paper have the equal rate $\lambda_{ON}$ and $\lambda_{OFF}$, and they have the same bandwidth.

\subsection{Collaborative Sensing Model}
We consider the CR network consisting of the secondary users which cooperatively exploit the unused spectrum resources.
We adopt the distributed collaborative sensing model, in which all CR users first sense the spectrum individually, and then share the sensing reports with each other, and finally they locally make the decision on the spectrum availability by combining the received sensing reports\cite{A}. In this paper, we consider $N$ CR users in networks, belonging to the set $\{SE_1,SE_2,\cdots,SE_N\}$, and we assume any two users can exchange the reports through one hop or multi-hop manner.

We assume that each user is equipped with energy detectors, and the detection is executed in Rayleigh fading environment.  The choice of energy detection is due to its widespread acceptance and ease of implementation and analysis. During the cooperation, the secondary user will share their binary decisions over these M channels\cite{hanzhu}, and we denote sensing report of the secondary user $SE_x$ as $\mathbf{R^x}=[R_1^x, R_2^x,\cdots,R_M^x ]$, in which $R_i^x=1$ denotes the channel $\widetilde{C}_i$ is detected in the ON period, while $R_i^x=0$ denotes this channel is detected in the OFF period. In addition, we assume each secondary user adopts $k$ out of $n$ rule to combine the received reports.

\subsection{Entropy Selfishness Model}


In collaborative spectrum sensing, some selfish users may claim the sensing results from other users as the fresh ones without scanning the spectrum.
In particular, we consider the following two kinds of selfish behaviors.
\begin{itemize}
  \item{\textbf{Exhaustive Entropy Selfishness}} (EES): The selfish secondary user does not sense the spectrum at all during one sensing round. Instead, it will multicast one sensing report from others (or the slightly modified version) as its own. In collaborative sensing, it is challenging to distinguish an entropy selfish user from an honest one.
  \item{\textbf{Partial Entropy Selfishness}} (PES): The selfish secondary user aims to learn the spectrum states in a most cost-effective way: sensing a part of the spectrum during a cooperative round. In our proposed YouSense, the selfish user may try to recover the whole sensing report by only sensing a small proportion of the spectrum.
\end{itemize}
In this study, we do not consider the malicious attacks such as reporting inaccurate or even fake messages. In other words, the users considered in this paper want to maximize their benefits by selfish behavior, and they are not malicious to deliberately interrupt the system. We believe malicious attacks deserve separate studies and there are existing works like\cite{shuai,hanzhu,NDSS11,shuai2}, which have proposed a series of countermeasures to prevent these attacks.
Besides, we don't consider the traditional selfishness that could be addressed by the existing mechanism,
including reputation based\cite{reputation1,reputation2}, credit based\cite{credit2,credit5}, or kind of cooperation strategies\cite{beibei,qianzhang}.
At last, the collusion attack, which is an open problem for wireless network security, is not considered in this study either.

\section{YouSense: a one-time pad based incentive design}
In this section, we will introduce the YouSense in details. The basic idea of YouSense is to construct a trapdoor one-time pad for each sensing report by combining the original report and a random key. Such a one-time pad based encryption could prevent the entropy selfish users from accessing the original sensing report while enabling the honest users to recover the report. Different from traditional OTP, in which the receivers use the encryption random key for plaintext recovery, the proposed scheme does not require the random key delivery process. Instead, YouSense proposes a novel way to enable the honest users to recover the pad as well as the original sensing report by exploiting a unique characteristic of collaborative spectrum sensing, which is the correlation among the sensing reports of different CR users. In particular, since different CR users sense the same channel set, it is obvious that they actually share some similar spectrum observations, or spectrum sensing reports. These shared common sensing reports could be exploited by honest CR users to recover the encryption key by known plaintext attack.

%
%

And yet, due to the incomplete plaintext in key recovery, the recovery successful rate at honest user side is found not high. We further deal with this problem by reducing the cardinality of set of the possible pads. In this way, the difficulty in pad recovery can be decreased, and therefore the honest CR user can recover the plaintext at a higher successful rate. Further, we discuss how to select this subset, not only ensuring the high success ratio, but also keeping YouSense secure against entropy selfish users. In the following, we will present the detailed design of YouSense.

\subsection{The Basic YouSense Design}
The basic YouSense design includes two separate parts. For the CR user who is going to multicast the results, he should encrypt the reports by OTP. And for the honest user who hopes to decode this ciphertext, he should execute the plaintext recovery process. Next, we will introduce them separately.

\subsubsection{Sensing Report Encryption at Sender Side}

YouSense requires each CR user to share OTP ciphertexts. Suppose the original sensing result of the user $SE_x$ is $\mathbf{R}^x=[R_1^x,R_2^x,\cdots, R_M^x]$, the encryption can be described:
\begin{equation}
    \mathbf{E}^x=\mathbf{R}^x \oplus \mathbf{K}^x
\end{equation}
where $\mathbf{K}^x=[K_1^x,K_2^x,\cdots, K_M^x]$ is the random key usually having the same length as the plaintext $\mathbf{R}^x$.

Though the entropy selfish users can receive this OTP ciphertext in collaborative sensing, they couldn't learn the original sensing results of the user $SE_x$, because the random key is only held by the sender and entropy selfish users cannot decode the ciphertext without the secret key. For honest CR users who have sensed the spectrum, they can recover this ciphertext by spectrum sensing. In the following, we will introduce the sensing report recovery method in details.

\subsubsection{Sensing Report Decryption by Spectrum Sensing}
\begin{figure}
\centering
\includegraphics[width=0.42\textwidth]{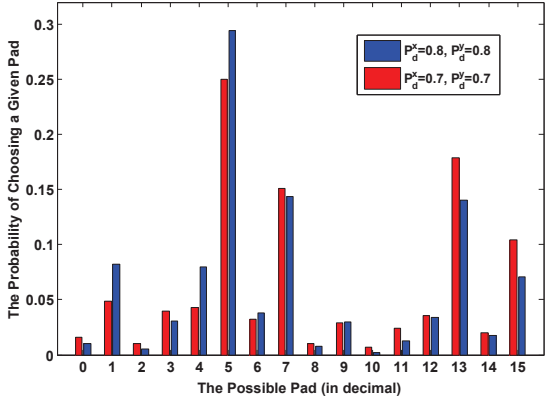}
\vspace{-0.3cm}
\caption{\textbf{The Evaluation of Pad Recovery.} In our simulation, we assume $M$=$4$, $\widetilde{C}_1$=$\widetilde{C}_3$=$1$, and $\widetilde{C}_2$=$\widetilde{C}_4$=$0$. The target false positive rate is set to be 0.1, and without lose of generality, we suppose the detection rate $P_d^x$ (and $P_d^y$) over each channel is equal. We also initiate the pad $\textbf{K}^x=[0,1,0,1]$.}
\vspace{-0.5cm}
\end{figure}
Spectrum sensing enables the CR user to recover the OTP key by known plaintext attack. The plaintext is learned by the receiver via examining his own sensing results. Suppose the CR user $SE_y$ receives the ciphertext $\textbf{E}^x$, and has obtained his sensing results $\mathbf{R}^y=[R_1^y,R_2^y,\cdots,R^y_{M}]$ by spectrum sensing. Then, the recovery process is actually to pick out the most probable pad from set $\mathbb{K}=\{\textbf{K}_1,\textbf{K}_2,\cdots,\textbf{K}_{2^M}\}$ which contains all the possible pads.
We first calculate the probability that a given pad $\mathbf{{K}}_j=[K_{j,1},K_{j,2},\cdots,K_{j,M}]$ is used by sensing report sender:
\begin{equation}
    Pr(\textbf{K}^x=\mathbf{{K}}_j)=\prod_{i\in\{1,2,\cdots,M\}} {Pr(K_i^x={K}_{j,i})}
\end{equation}
For each bit $K^x_i$, if $R^x_i=R^y_i$, then $K^x_i=R^y_i\oplus E_i^x$, otherwise $K^x_i$ will take the other value. Thus, we get:
\begin{equation}
    \begin{split}
        & Pr(K_i^x={K}_{j,i})=(E^x_i\oplus R^y_i\oplus {K}_{j,i})Pr(R_i^x \neq R_i^y)+ \\
        & \ \ \ \ \ \ \ \ \ \ \ \ \ \ (E_i^x\oplus R_i^y\odot{K}_{j,i})Pr(R_i^x = R_i^y)
    \end{split}
\end{equation}
the probability of $R_i^x$=$R_i^y$ reveals how much the receiver can learn the plaintext by examining his own reports, and it's usually larger than 0.5 due to detection correlation\cite{cosensing1}. Then, for each bit $K^x_i$, it's more likely to be $E^x_i\oplus R^y_i$. Then we get:
\begin{equation}
    Pr(\textbf{K}^x=\textbf{E}^x\oplus \textbf{R}^y)=\max_{\textbf{K}_j\in \mathbb{K}}{Pr(\textbf{K}^x=\textbf{K}_j)}
\end{equation}
This result reveals that the pad $\textbf{E}^x\oplus \textbf{R}^y$ has the largest probability to be the correct one, thus it should be selected as the pad recovery result. However, since the sensing results of different CR users are not always equal (that's why collaborative sensing makes sense), our pad cracking method may sometimes fail. This problem is shown in Fig. 2.
It's clear that though the right pad has the highest probability to be selected, the pad recovery successful rate is still unsatisfying. This makes the basic design less practical. We will introduce an advanced design in the following section.


\subsection{The Advanced YouSense Design}
In basic design, the primary reason for the low successful rate is that the different CR users may have a few different sensing reports due to spatial diversity, and the recipient cannot learn all the plaintext bits by checking his own reports. This incomplete plaintext information incurs limited pad recovery capacity. In order to address this problem, the advanced YouSense design reduces the recipient's difficulty in pad recovery by decreasing the possible pad set, while still ensuring the secrecy towards the entropy selfish users.


In advanced YouSense, a subset of $\mathbb{K}$ is predefined as a possible pad set for OTP encryption, and it is openly known by all the secondary users. For ease of presentation, we denote this subset as $\mathbb{\overline{K}}=\{{\overline{\textbf{K}}_1},{\overline{\textbf{K}}_2},\cdots,{\overline{\textbf{K}}_m}\}$. In collaborative spectrum sensing, the report sender randomly selects one pad from $\mathbb {\overline{K}}$ to create the encrypted sensing reports, and thus the honest CR user just needs to search the set $\mathbb {\overline{K}}$ instead of $\mathbb {{K}}$ for the correct pad. According to previous analysis, when the CR user $SE_y$ receives the ciphertext $\textbf{E}^x$ from user $SE_x$, $R_i^y\oplus E_i^x$ is more likely to be the correct pad bit for user $SE_y$. Thus, the pads in $\mathbb{\overline{K}}$ with $R_i^y\oplus E_i^x$ as the $i$-th bit is weighted in the pad recovery process. For the simplicity of presentation, we maintain a vector $\textbf{W}=\{w_1,w_2,\cdots,w_m\}$ to record each pad's overall weight, and finally the pad with the highest weight will be selected as pad recovery result. This advanced design is summarized as the following algorithm:
\vspace{-0.2cm}
\begin{algorithm}[htbp]
\caption{The Advanced YouSense Design} \label{advanced}
\begin{algorithmic}[1]
    \STATE{\emph{Procedure} \textbf{Secret}\_\textbf{Gen}($\mathbf{R}^x$, $\mathbb{\overline{K}}$)}\ \  // Report Encryption \ \
    \FOR{Each collaborative sensing round}
        \STATE{Sense the spectrum and obtain the vector $\mathbf{R}^x$}
        \STATE{Randomly select a pad $\mathbf{\overline{K}}_j$ from set $\mathbb{\overline{K}}$}
        \STATE{Generates $\mathbf{E}^x=\mathbf{R}^x \oplus \mathbf{\overline{K}}_j$}
        \STATE{Declare the one-time pad ciphertext $\mathbf{E}^x$}
    \ENDFOR \ \ \ // Each user executes this procedure \ \ \ \ \ \ \ \ \ \ \ \ \ \ \ \ \ \ \ \ \ \ \ \ \ \ \ \ \ \ \ \ \ \ \ \ \ \ \ \ \ \ \ \ \ \ \ \ \ \ \ \ \ \ \ \ \ \ \ \ \ \ \ \ \ \ \ \ \ \ \ \ \ \ \ \ \ \ \ \ \ \ \ \ \ \ \ \ \ \ \ \ \ \ \ \
    \STATE{\emph{Procedure} \textbf{Secret}\_\textbf{Crack}($\mathbf{R}^y$,$\mathbf{E}^x$,$\mathbb{\overline{K}}$)} // Pad recovery
    \FOR{Each collaborative sensing round}
        \STATE{Scan the spectrum and obtain $\mathbf{R}^y$}
        \STATE{Receive the one-time pad ciphertext $\mathbf{E}^x$}
        \STATE{Initiate the weight $W=[w_1,w_2,\cdots,w_m]=0$}
        \FOR{Each bit $R_i^y\oplus E_i^x$ in $\mathbf{R}^y\oplus \mathbf{E}^x$}
            \FOR{Each possible pad $\mathbf{\overline{K}}_j$ in $\mathbb{\overline{K}}$}
                \IF{$R_i^y\oplus E_i^x==\mathbf{\overline{K}}_{j,i}$}
                    \STATE{$w_j=w_j+1$}
                \ENDIF \ \ \ \ \ // Or the weight goes unchanged
            \ENDFOR
        \ENDFOR  \ \ \ \ \ \ \   // Calculates the weights of pads
        \STATE{$w_{h'}=\max_{\{w_h\in \mathbf{W}\}}{w_h}$, $1 \leq h' \leq m$}
        \STATE{\textbf{return}\ \ $\mathbf{\overline{K}}_{h'}$}
    \ENDFOR
%
%
\end{algorithmic}
\end{algorithm}
\vspace{-0.2cm}

The only remaining issue is how to select the proper subset $\mathbb{\overline{K}}$.
we will discuss the subset design in details.

\subsubsection{The Subset Design For EES}
In advanced YouSense Design, the user selects a random pad from $\mathbb{\overline{K}}$ to encrypt the sensing reports. Though adoption of the subset means that the OTP is no longer unconditional secure, this design can still prevent EES users from learning the sensing reports, as long as the pads are properly selected. In the following, we will discuss how to select the secure pads.
Before describing the detailed design method, we first introduce the definition of \emph{sensing report masking level} to measure how much the known ciphertext bits reduce the EES user's uncertainty about the relevant channel state. By referring to the concept of mutual information\cite{information}, we have the following definition:

\textbf{DEFINITION 1:} \emph{(Sensing Report Masking Level) Let $E^x_i$ denote $SE_x$'s ciphertext bit about channel $\widetilde{C}_i$. During a collaborative sensing process, the entropy selfish user $SE_y$ observes $E^x_i$, and predicts the possible channel state of $\widetilde{C}_i$. We define $E^x_i$'s sensing report masking level $I(\widetilde{C}_i, E_i^x)$ as:}
\begin{equation}\label{}
    \begin{split}
    & \ \ \ \ \ \ \ I(\widetilde{C}_i, E_i^x)  \\
    & =\sum\limits_{\widetilde{C}_i\in\{0,1\}} \sum\limits_{E_i^x\in\{0,1\}}{p(\widetilde{C}_i,E_i^x) log(\frac {p(\widetilde{C}_i,E_i^x)}{p_1(\widetilde{C}_i) p_2(E_i^x)}) }
    \end{split}
\end{equation}
\emph{where $p_1(\widetilde{C}_i)$ and $p_2(E_i^x)$ are the marginal probability distribution functions of $\widetilde{C}_i$ and $E_i^x$.}

From the definition, we can find the report masking level is non-negative, and this metric is proportional to the information that the PES users learn from the ciphertext bits. Next, we will present our subset design whose report masking level equals to zero, which shows each bit of the OTP ciphertext gives no information about the relevant channel availability decision.

In our design, $\mathbb{\overline{K}}$ is comprised of multiple \emph{secure pad pairs}. We coin two randomly generated pads $\mathbf{\overline{K}}_j, \mathbf{\overline{K}}_{j'}$ as secure pad pair, if this two pads are complementary. In other words, these two pads satisfy the equation $\mathbf{\overline{K}}_j\odot \mathbf{\overline{K}}_{j'}=\mathbf{0}$. We prove in the following theorems that the subset consisting of these secure pairs can get the sensing report masking level to be zero:

\emph{\textbf{Theorem 1:} In YouSense, for every pad $\mathbf{\overline{K}}_j\in \mathbb{\overline{K}}$, if there exists a pad $\overline{K}_{j'}\in \mathbb{\overline{K}}$ that satisfies $\mathbf{\overline{K}}_j\odot \mathbf{\overline{K}}_{j'}=\mathbf{0}$, then we have $E_i^x$ sensing report masking level:}
\begin{equation}\label{}
    I(\widetilde{C}_i, E_i^x)=0
\end{equation}

\emph{\textbf{Proof:}} See Appendix A.

This theorem means that the subset $\mathbb{\overline{K}}$ selected by the above method enables the ciphertext bit $E_i^x$ to leak no state information in collaborative sensing. And yet, the EES user actually receives multiple secrets $\mathcal{S}_i=\{E_i^1,E_i^2,\cdots,E_i^N\}$ from different users, we should further investigate the sensing report masking level of the set $\mathcal{S}_i$. Then we have:

\emph{\textbf{Theorem 2:} If the selected subset $\mathbb{\overline{K}}$ makes $I(\widetilde{C}_i, E_i^x)=0, i\in\{1,2,\cdots,M\}$, then the sensing report masking level of the set $\mathcal{S}_i$ is:}
\begin{equation}\label{}
    I(\widetilde{C}_i, \mathcal{S}_i)=0
\end{equation}

\emph{\textbf{Proof:}} $I(\widetilde{C}_i,E_i^x)=0$ demonstrates the channel $\widetilde{C}_i$'s state is independent with $E_i^x,\forall x \in\{1,2,\cdots,N\}$. In addition, since secrets are generated independently, $E_i^x,\forall x \in \{1,2,\cdots,N\}$ is independent with $E_i^l,\forall l \in \{1,2,\cdots,N\}/x$. Thus, all elements in set $\mathcal{S}_i\cup \widetilde{C}_i$ are mutually independent with each other, and therefore we could have following equation:
\begin{equation*}
    p(\widetilde{C}_i,\mathcal{S}_i)=p(\widetilde{C}_i)p(\mathcal{S}_i)
\end{equation*}
Referring the equation (5), we can obtain the theorem.
\ \ \ \ \ \ $\blacksquare$

The above discussion reveals that if the subset $\mathbb{\overline{K}}$ in YouSense contains only the secure pairs, no matter with single or multiple OTP ciphertexts, EES users cannot derive any useful information from others' sensing results.
This result also shows whether the YouSense design can defense against EES just lies in whether the subset includes only the secure pairs, with no relation with the pair numbers. Since the difficulty in pad recovery is proportional to the cardinality of the subset, we can simply adopt only one pair in $\mathbb{\overline{K}}$ to facilitate the honest user's plaintext recovery process. In this situation, we can obtain the pad recovery successful rate (the probability that weight of the correct pad is larger than any other pad):

\emph{\textbf{Theorem 3:} When YouSense adopts $\mathbb{\overline{K}}=\{\mathbf{\overline{K}}_j,\mathbf{\overline{K}}_{j'}\}, \mathbf{\overline{K}}_j\odot \mathbf{\overline{K}}_{j'}=\mathbf{0}$, the recovery successful rate $P_s$ is:}
\begin{equation}
    P_s(M)=\sum\limits_{k=\lceil \frac{M}{2}\rceil}^{M}{\sum_{A\in F_k}{\prod_{i\in A}{\eta_i \prod_{l\in A^c}{(1-\eta_l)}}}}
\end{equation}
,where $\eta_i=\mathbb{P}(|R_i^x-R_i^y|=0)$, and $F_k$ is the set of all subsets of $k$ integers that can be selected from $\{1,2,\cdots,M\}$, $A^c$ is the complement of A.

\emph{\textbf{Proof:}} For each bit of the ciphertext $E_i^x$, if the sensing result $R_i^x=R_i^y$, the CR user $SE_y$ will successfully recover this bit, and the correct pad will be weighted by 1, and the weight of the other one will remain unchanged. Thus, if these two users share same observations towards at least half of the channels, the weight of the right pad will be bigger than the other one, that is to say, the pad is successfully recovered. Given the false positive/negative rate of secondary user $SE_x$ and $SE_y$ towards channel $\widetilde{C}_i$, we could obtain:
\begin{equation*}\label{}
    \begin{split}
        &\eta_i =\mathbb{P}(R_i^x=R_i^y|\widetilde{C}_i=-1)+\mathbb{P}(R_i^x=R_i^y|\widetilde{C}_i=1) \\
        & \ \  =(1-2p_f+2p_f^2)p_0+(2p_{m,i}^x p_{m,i}^y+1-p_{m,i}^x-p_{m,i}^y)p_1
    \end{split}
\end{equation*}
We suppose the radio channels are independent, and then we can get the cracking successful rate. \ \ \ \ \ \ \ \ \ \ \ \ \ \ \ \ \ \ \ \ \ \ \ \ \ \ \ \ \ \ \ $\blacksquare$

In the experiment, we can see the larger $M$ leads to larger pad recovery successful rate. Since the number of target channels in spectrum sensing is usually more than one, the parameter $M$ can be large enough to keep the successful rate quite high. This fact shows that our YouSense design makes little impact on collaborative spectrum sensing performance.

\subsubsection{The Subset Design For PES} When subset includes one secure pair, YouSense can effectively prevents EES users from learning the channel states in collaborative sensing. However, this subset selection method may also provide the attack opportunity for the PES users, which is shown in Fig.3.

Though only sensing parts of the spectrum, the PES user may still be able to crack the whole ciphertext by this limited channel information.
Suppose the target recovery successful rate is set as $p_{tar}$ and $\varphi=P_s^{-1}(p_{tar})<M$.
For ease of presentation, we divide the pad $\mathbf{K}$ into ${\lceil \frac{M}{\varphi}\rceil}$ blocks:
\begin{equation*}
\mathbf{K}=[\mathbf{K}\langle1\rangle,\cdots,\mathbf{K}\langle\lfloor \frac{M}{\varphi}\rfloor\rangle,\mathbf{K}\langle{\lceil \frac{M}{\varphi}\rceil}\rangle]
\end{equation*}
where $\mathbf{K}\langle i \rangle=[K_{1+\varphi*(i-1)},\cdots,K_{\varphi*i}],$ $i<{\lceil \frac{M}{\varphi}\rceil}$ having $\varphi$ entries, and $\mathbf{K}\langle{\lceil \frac{M}{\varphi}\rceil}\rangle=[K_{1+\varphi*{\lfloor \frac{M}{\varphi}\rfloor}},\cdots,K_{M}]$ including $M \bmod \varphi$ elements. When the subset $\mathbb{\overline{K}}$ contains just one pad pair $\{\mathbf{\overline{K}}_j,\mathbf{\overline{K}}_{j'}\}$, the PES user can achieve the target recovery successful rate by sensing only the spectrum corresponding to any pad block (except the last block).
This attack holds just because the subset consisting of only one pad pair is somewhat easy to crack when the ciphertext has multiple bits. The correlation between the pad blocks can help the PES user crack the whole ciphertext by decoding any pad blocks. For example, if PES users crack the first pad block and learn $\mathbf{\overline{K}}_{j'}\langle 1 \rangle$ is correct, they can simply deduce the value of other pad blocks.

To address PES, we propose the subset design for PES. The basic idea is to break down the connection between the pad blocks by adding more secure pairs into $\mathbb{\overline{K}}$. In another word, we hope our subset design can achieve the following property:
\begin{equation}
    Pr(\mathbf{K}^x\langle l \rangle\ | \ \mathbf{K}^x\langle q \rangle)=Pr(\mathbf{K}^x\langle l \rangle), \ l\neq q
\end{equation}
We implement this idea by interweaving the basic pad pair $\{\mathbf{\overline{K}}_j,\mathbf{\overline{K}}_{j'}\}$. For each pad block $\mathbf{K}^x\langle l \rangle$, the sender $SE_x$ randomly selects $\mathbf{\overline{K}}_j\langle l \rangle$ or $\mathbf{\overline{K}}_{j'}\langle l \rangle$ as the encryption pad block. By this way, each pad block $\mathbf{K}^x\langle l \rangle$ is independent with any other block $\mathbf{K}^x\langle q \rangle$. Then, the adopted subset is actually $\mathbb{\overline{K}}=\{\mathbf{K}\ |\ \forall l, \mathbf{K}\langle l \rangle\in\{\mathbf{\overline{K}}_j\langle l \rangle,\mathbf{\overline{K}}_{j'}\langle l \rangle\}\ \}$. The detailed algorithm can be described in Algorithm 2. We can notice when $\varphi=M$, the algorithm is just the subset design for EES.
\vspace{-0.5cm}
\begin{algorithm}[htbp]
\caption{The Subset Design in YouSense} \label{basic}
\begin{algorithmic}[1]
    \STATE{\emph{Procedure} \textbf{Subset}\_\textbf{Gen}($M$)}\ \ \  // Subset Generation \ \
    \STATE{Initialize the parameter $\varphi=P_s^{-1}(p_{tar})$}
    \STATE{Generate $M$ random binary bits as $\mathbf{\overline{K}}_j$}
    \STATE{Obtain $\mathbf{\overline{K}}_{j'}=\mathbf{1}\oplus\mathbf{\overline{K}}_j$} \ \ \ //Generate a secure pair
    \STATE{Initialize $\mathbb{\overline{K}}=\{\mathbf{\overline{K}}_j,\mathbf{\overline{K}}_{j'}\}$}
    \FOR{each pad block $\mathbf{K}\langle l \rangle, l=1,2,\cdots,\lceil \frac{M}{\varphi}\rceil$}
        \STATE{Initialize a temporary set $T=\mathbf{0}$}
        \FOR{each entry $\mathbf{\overline{K}}_\ast$ in the subset $\mathbb{\overline{K}}$}
            \STATE{Initialize $\mathbf{\overline{K}}_{new}=\mathbf{\overline{K}}_\ast$}
            \STATE{$\mathbf{\overline{K}}_{new}\langle l \rangle=\mathbf{1}\oplus \mathbf{\overline{K}}_{new}\langle l \rangle$}
            \STATE{Add the pad $\mathbf{\overline{K}}_{new}$ to set T}
        \ENDFOR
        \STATE{Add all elements in set T to the subset $\mathbb{\overline{K}}$}
    \ENDFOR
    \STATE{Broadcast the subset $\mathbb{\overline{K}}$ to all CR users}
\end{algorithmic}
\end{algorithm}
\vspace{-0.2cm}

If $M$ can be divided exactly by $\varphi$, the recovery successful rate can reach the target probability $p_{tar}$, because there is at least $\varphi$ bit difference between any two possible pads in $\mathbb{\overline{K}}$. In other cases, we can also achieve this target rate by transforming $\mathbf{R}^x$ to $[\mathbf{R}^x,R^x_1,\cdots,R^x_{\lceil{\frac{M}{\varphi}}\rceil*\varphi-M }]$ and executing subset generation algorithm with parameter $\varphi*\lceil \frac{M}{\varphi}\rceil$.

\begin{figure}
\centering
\includegraphics[width=0.38\textwidth]{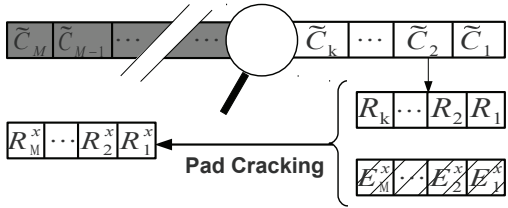}
\vspace{-0.2cm}
\caption{\textbf{Partial Entropy Selfishness:} in PES, the entropy selfish users try to crack the pad by sensing parts of the spectrum.}
\vspace{-0.5cm}
\end{figure}

\begin{figure*}[htbp]
  \centering
  \subfigure[the Number of Secure Pairs is 1]{\includegraphics[width=0.242\textwidth]{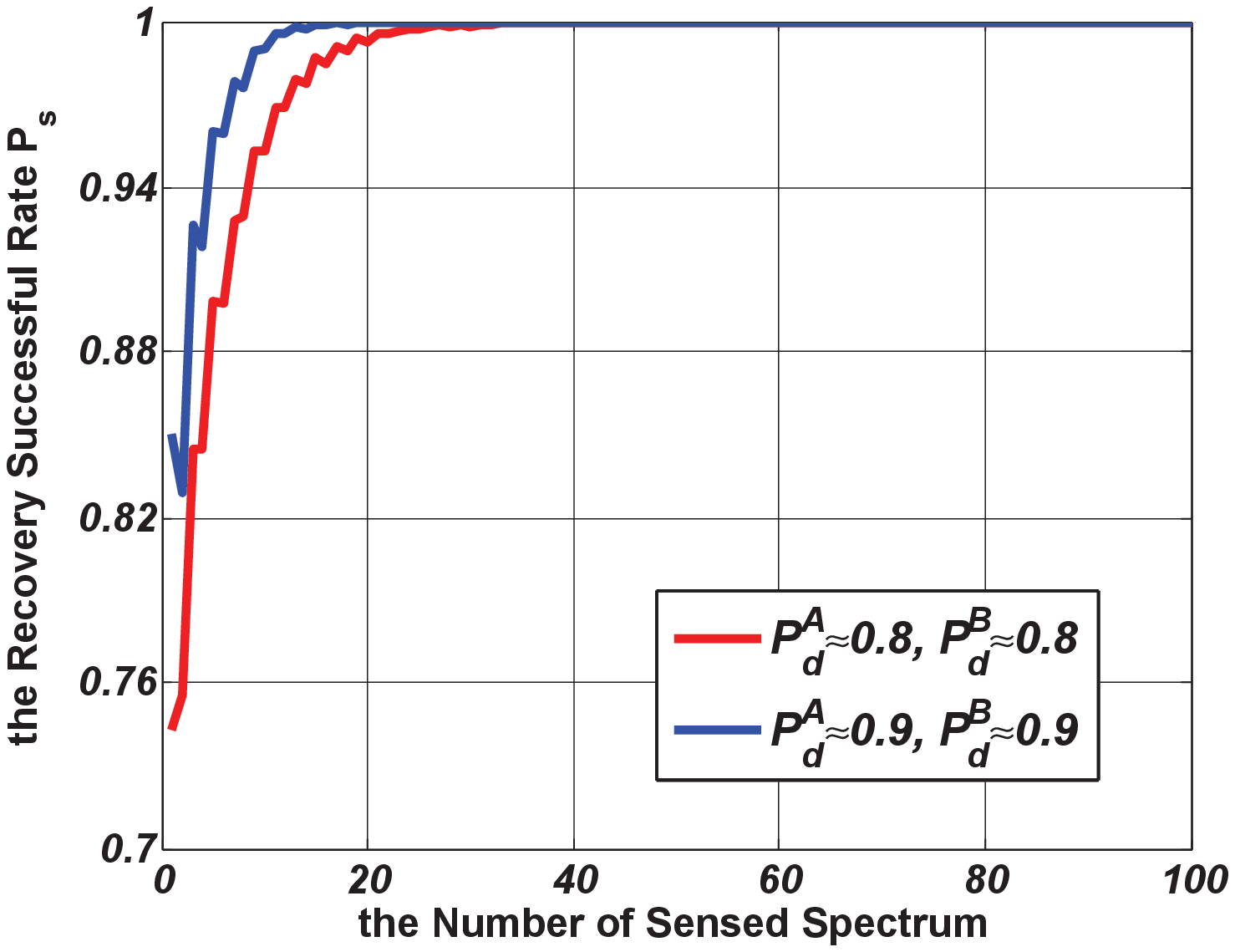}}
  \subfigure[the Number of Secure Pairs is 2]{\includegraphics[width=0.24\textwidth]{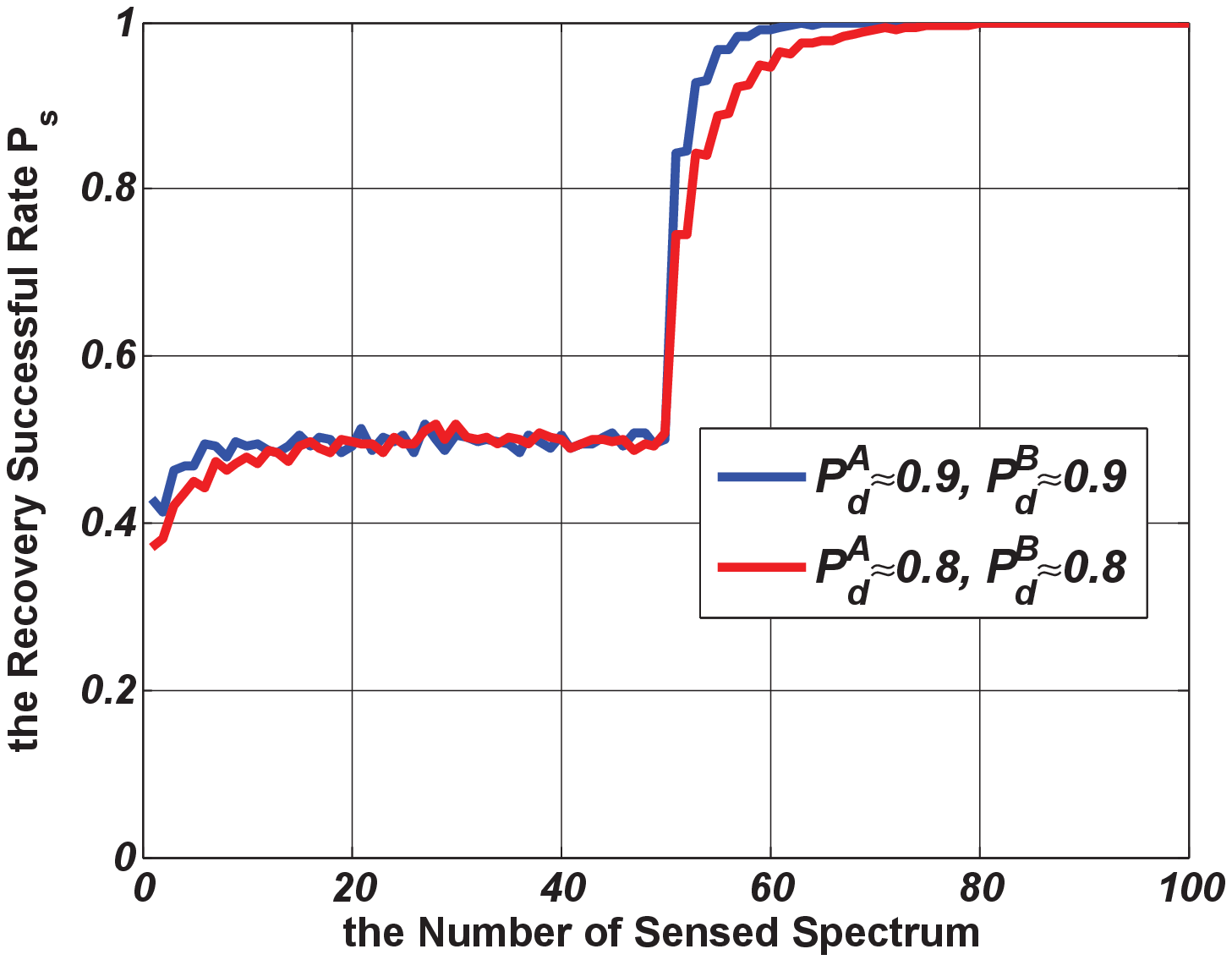}}
  \subfigure[the Number of Secure Pairs is 4]{\includegraphics[width=0.242\textwidth]{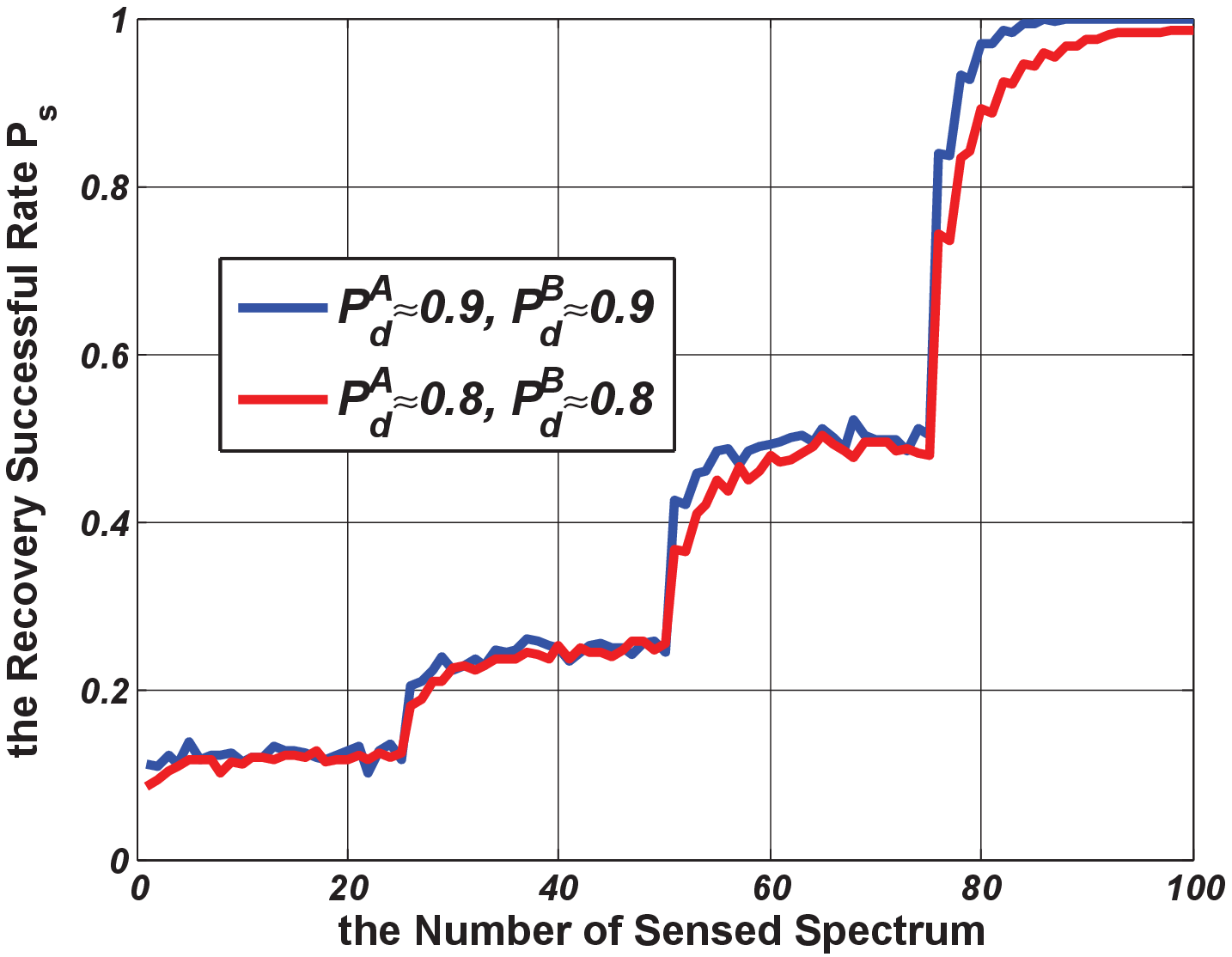}}
  \subfigure[the Number of Secure Pairs is 10]{\includegraphics[width=0.24\textwidth]{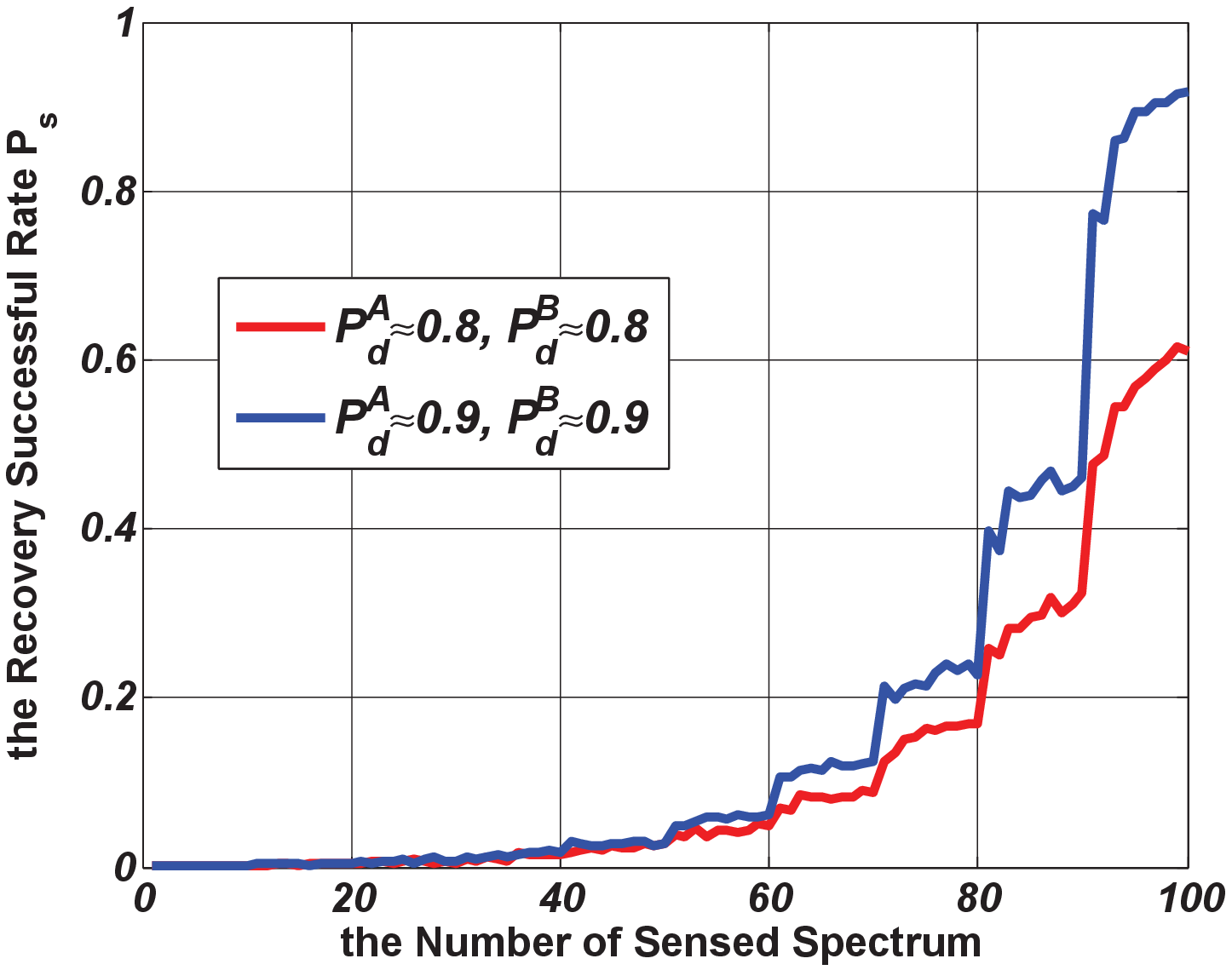}}
  \vspace{-0.1cm}
  \caption{\textbf{the Recovery Successful Rate} (In the simulation, we adopt Majority-Rule as the report combination rule.) }\vspace{-0.4cm}
\end{figure*}
\subsubsection{The Impact of History Based Attack}

It is possible the selfish users choose to scan the spectrum in one time slot, and launches entropy selfishness in next one. Based on the previous sensing results, the users may also be able to crack the OTP ciphertext.
The effectiveness of history-based attack largely depends upon the correlation degree between the sensing results of two continuous time slots. If the correlation are relatively high, in another word, the sensing results vary slowly, this attack will be effective. Here, we argue that this situation is not likely to happen, because it indicates the low efficiency of collaborative sensing, and the sensing frequency is often reduced to achieve higher data throughput\cite{tradeoff1}, \cite{tradeoff2}. In addition, even if previous sensing results is able to help crack the ciphertext, our YouSense design is still able to prevent entropy selfishness. This history information can not only help the entropy selfish user to crack the ciphertext, but also improve the honest users' sensing performance when combined with spectrum sensing results. Thus, there is still sensing performance gap between the honest users and the entropy selfish ones,   and we can defense against this kind of history based attack by simply reducing the parameter $\varphi$.

\subsubsection{The Impact of Shadowing on YouSense}
Shadowing occurs when the large obstruction obscures the main signal path between the primary user and the CR user,
and according to\cite{shadowco}, shadowing correlation displays distance dependence. Thus, the sensing results of the physically adjacent users will have higher correlation, while those of the users far from each other would have lower correlation. To get honest users to reliably recover the pad, we adopt larger parameter $\Omega \varphi$ when YouSense is applied in shadowing, where $\Omega>1$ is the prolong parameter and is determined by the environment.
This change will not incur insecurity to the system. Though the selfish user can successfully launch PES when he gets the ciphertext sent by near users, this user can't crack the ciphertext of far users unless he senses the spectrum as required.

\section{Performance Evaluation}
We evaluate the effectiveness and the efficiency of YouSense in terms of
the following aspects: 1) Setup of our experiments; 2) The Impact of Entropy Selfishness on Distributed Spectrum Sensing;
3) The Impact of YouSense on Collaborative Sensing; and 4) The Effectiveness of the YouSense.

%
%

\subsection{Simulation Setup}
In the simulation, we adopt the Universal Software Radio Peripheral (USRP) to evaluate our scheme. A USRP is utilized as the transmitter to simulate the primary users. We also use other USRPs as the receivers to sense the spectrum periodically. All these USRPs are connected individually with the identical computers, which are Dell OptiPlex 760.

We adopt RFX2400 as the daughter board which can operate in range $2.3G\sim 2.7G$, and the transmitter modulation is APSK. We randomly select $2.58$Ghz, $2.6$Ghz, $2.62$Ghz to conduct our experiment. The receivers utilize the energy detection method, and the detectors' thresholds are determined by letting the false positive rate be $0.1$. We scan 4M bandwidth for each time, and we set the number of FFT bins, decimation to be $256$, $16$, respectively, and the gain to be $45$dB.

The receivers are set to sense synchronously, and the sensing period as well as the sensing time are set to be $10$ ms and $1$ ms, respectively. Here, the synchronization is realized via utilizing the computer clocks, both of which are synchronized in advance with the NTP server: s1a.time.edu.cn.

\subsection{The Impact of Entropy Selfishness on collaborative Sensing}
In this section, we investigate the influence of entropy selfishness on collaborative spectrum sensing. In the experiment, we set the the total number of secondary users joining collaborative sensing as $5$, and the number of honest users varies from $1$ to $5$. If the system doesn't take any countermeasure, an entropy selfish user will claim others' sensing results as his own and then use it to trade more sensing results. We conduct our experiment in low and high SNR situation, where the users' detection rate is approximately 0.8 and 0.9, respectively. We also do our experiment on 2.58Ghz, 2.6Ghz, 2.62Ghz separately, and obtain the average false positive/negative rate, which is often adopted to measure the collaborative sensing performance\cite{hanzhu,shuai}. The results are shown in Table 1. We can see along with the increase of the entropy selfish users, the false positive/negative rate of the collaborative sensing will rise significantly. When there is only 1 honest secondary user, the collaborative sensing will degrade to the individual sensing. Thus, thwarting entropy selfishness is of significance.

\begin{figure*}
  \centering
  \subfigure[False positive rate with one secure pair]{\includegraphics[angle=0,width=0.435\textwidth]{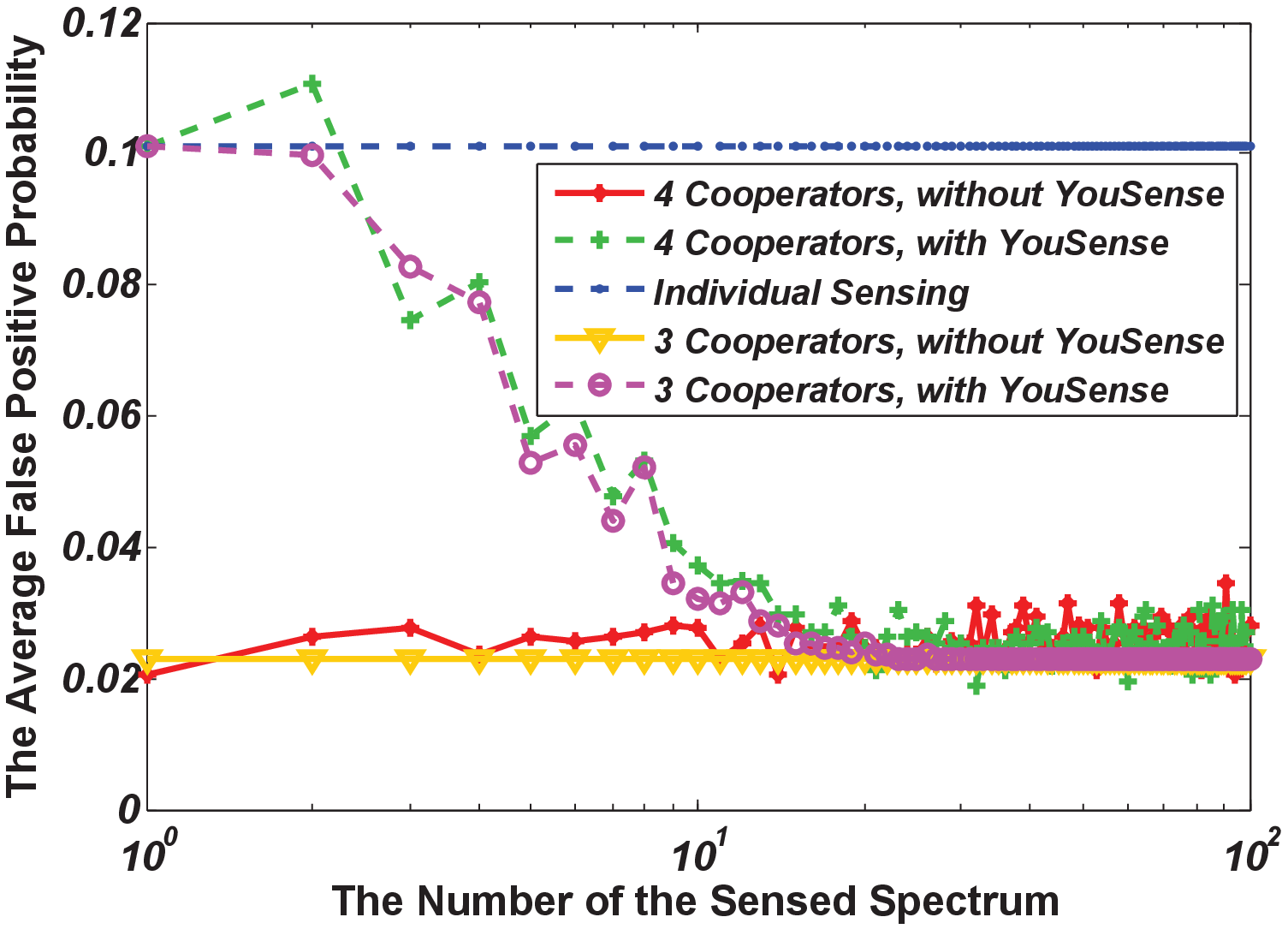}}
  \subfigure[False negative rate with one secure pair]{\includegraphics[angle=0,width=0.43\textwidth]{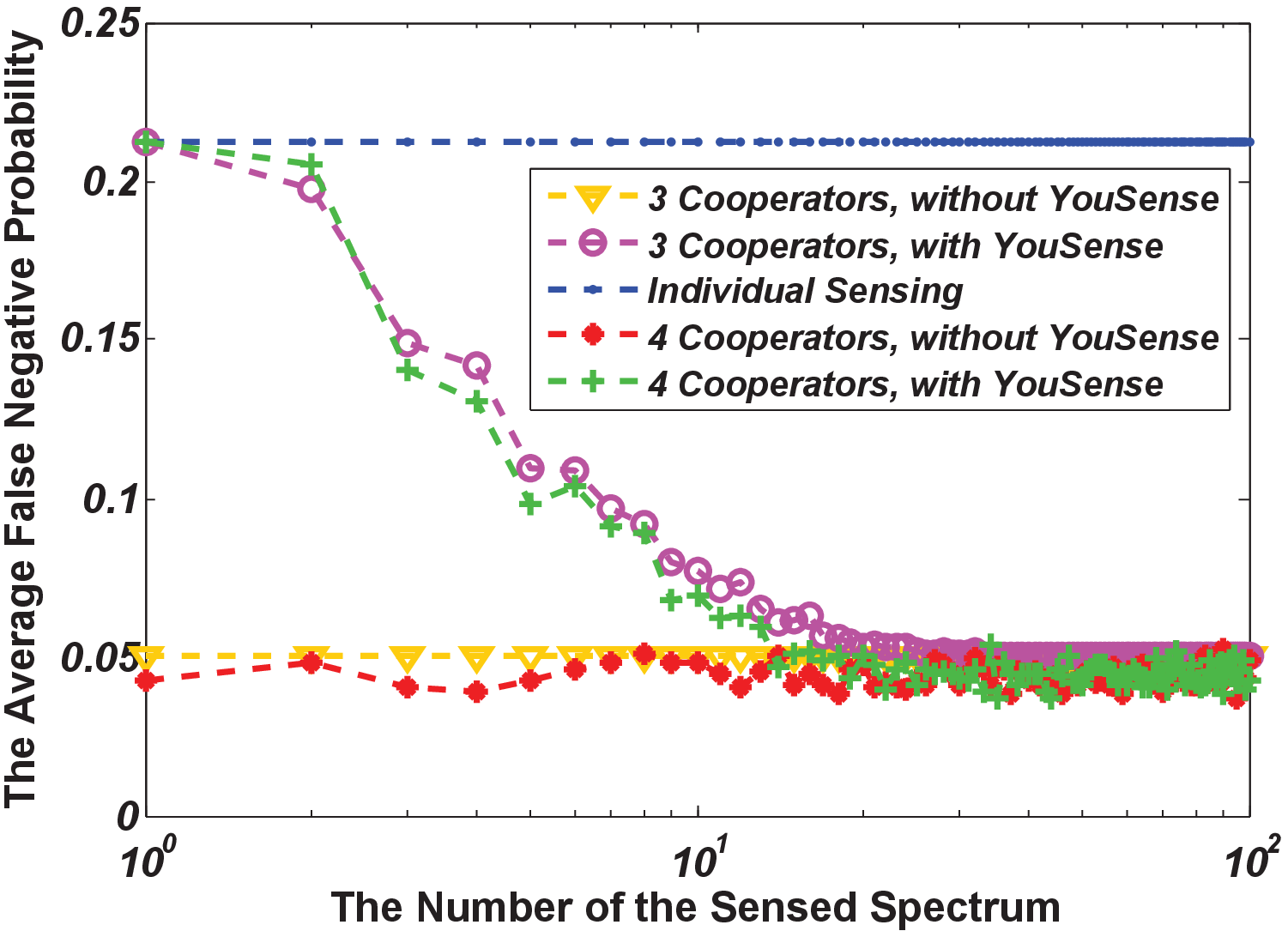}}
  \subfigure[False positive rate with multiple pairs (4 Cooperators)]{\includegraphics[width=0.43\textwidth]{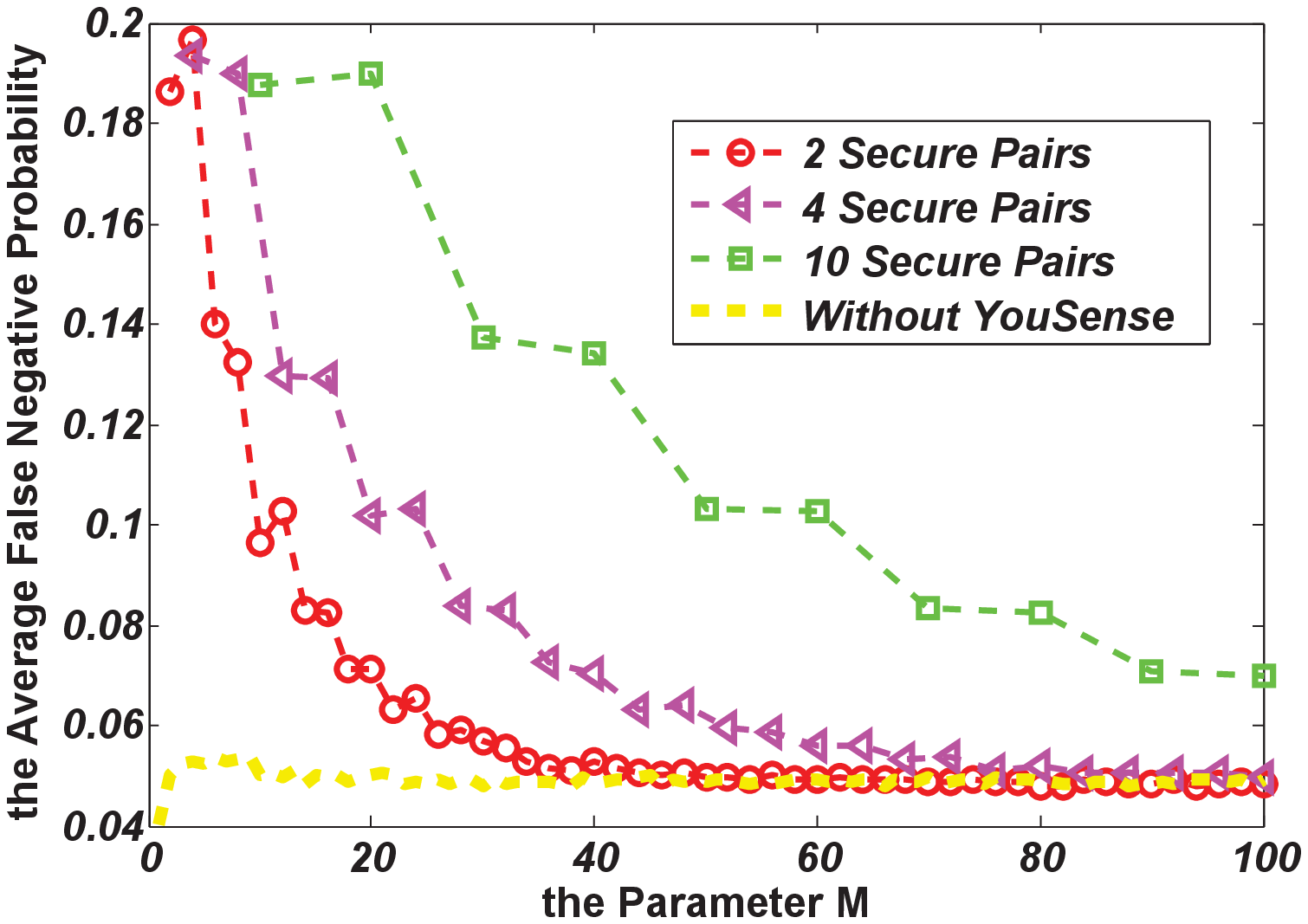}}
  \subfigure[False negative rate with multiple pairs (4 Cooperators)]{\includegraphics[angle=0,width=0.43\textwidth]{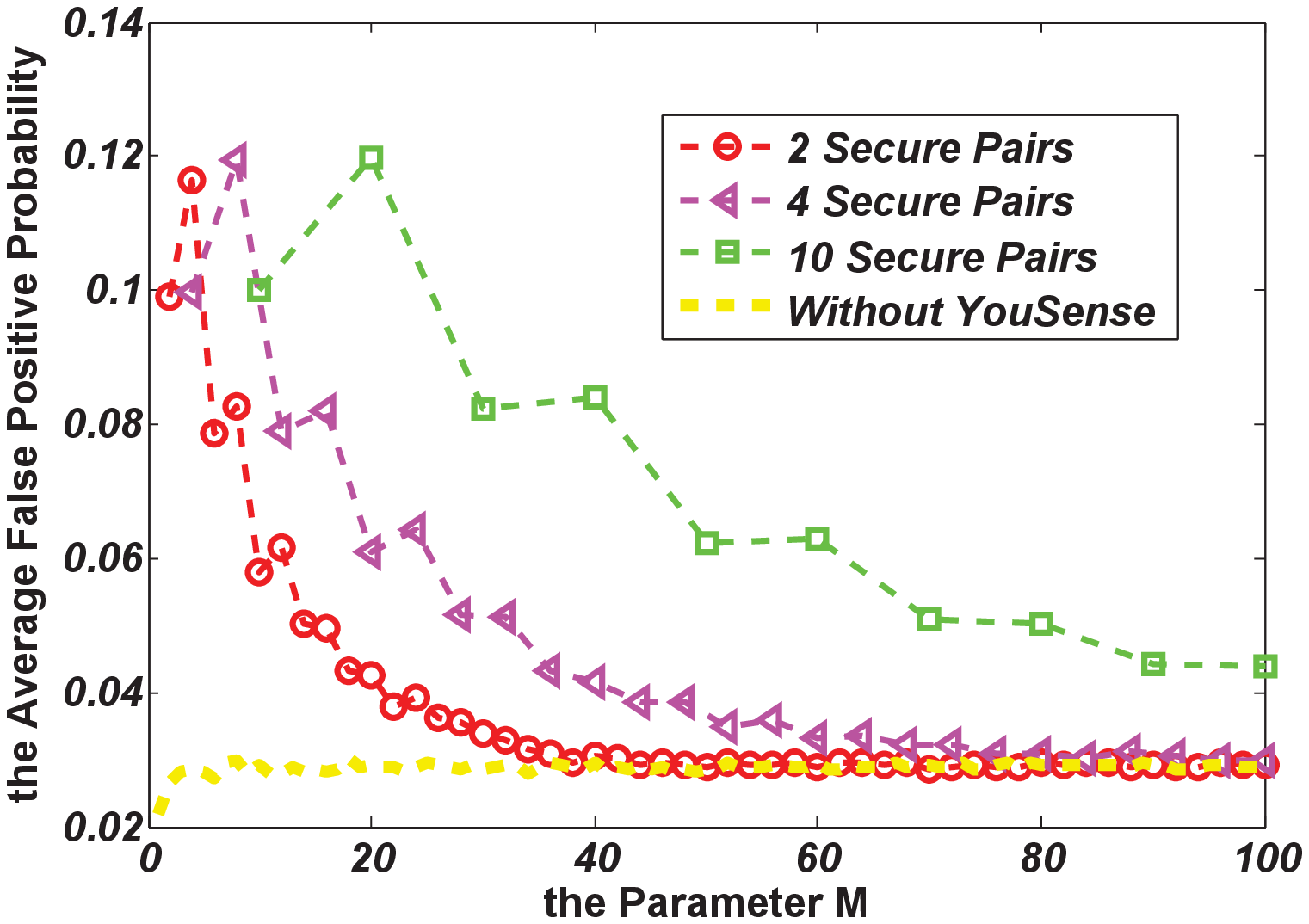}}
  \vspace{-0.1cm}
  \caption{\textbf{YouSense's Impact on Collaborative Spectrum Sensing Performance} (In the simulation, the influence of the YouSense design on collaborative spectrum sensing is measured by the metric of false positive/negative probability. A secondary user is selected as the sensing reports recipients, who recovers the OTP plaintexts and combines all sensing reports by the Majority-Rule.) }\vspace{-0.5cm}
\end{figure*}

\begin{table}
\renewcommand{\arraystretch}{0.8}
\begin{center}
\caption{Entropy Selfishness's impact on Collaborative Sensing}
\begin{tabular}{@{}cccccccc@{}}
\toprule Entropy selfish user No. &0&1&2&3&4 \\
\midrule False Positive Rate, high$\bar{\gamma}$ &0.019&0.052&0.081&0.147&0.138 \\
\midrule False Positive Rate, low$\bar{\gamma}$ &0.058&0.100&0.152&0.218&0.208 \\
\midrule False Negative Rate &0.008&0.033&0.051&0.108&0.096 \\
\bottomrule
\end{tabular}
\end{center}
\indent *\ The number of cooperators is set to be 5, and the number of entropy selfish users varies from 0 to 4. We find in low $\bar{\gamma}$ condition, the CR users' detection probability approximates 0.8, while in high $\bar{\gamma}$, this rate approximates 0.9.
\vspace{-0.5cm}
\end{table}

\subsection{The Impact of YouSense on Collaborative Sensing}
In this section, we evaluate YouSense's impact on the collaborative sensing performance. Let $M=100$, and all channels follow Poisson Process, and their state transition is independent. Without loss of generality, all these channels get parameter $\lambda_{OFF}=50$, $\lambda_{ON}=50$. For each channel $\widetilde{C}_i$, we randomly set it as 2.58Ghz, 2.6Ghz, 2.62Ghz.

We conduct our experiment in the following way: firstly, we do the experiment under channel's ON and OFF state individually, and obtain the sensing result samples of $SE_x$ and $SE_y$ in the corresponding state; and then the computer randomly selects a pair of sensing results from the corresponding sample sets as their sensing results. In following experiments, we also use this method to simulate the Poisson Process channel.

We firstly answer the question whether an honest CR user can accurately recover the OTP key. The subset $\mathbb{\overline{K}}$ is generated by the algorithm 2, and the number of secure pairs in $\mathbb{\overline{K}}$ takes the value of 1, 2, 4, and 10, separately. The simulation results are shown in Fig. 4. The x-axis represents the total number of the channels having been sensed by the recipient, and y-axis represents the pad recovery successful rate $P_s$. In Fig. 4(a), one secure pair is adopted, and the CR user needs to sense nearly 20 channels to safely decode the ciphertext. And in Fig. 4(b) and Fig. 4(c), with more secure pairs in $\mathbb{\overline{K}}$, the CR users should sense much more radio channels to achieve the similar pad recovery successful rate. This means the difficulty in pad recovery is increased with more pad pairs added. In Fig. 4(d), we can see when the subset contains ten pairs, the recovery successful rate may be unsatisfying even the CR user senses all the spectrum. This result reveals when the system takes a proper parameter $M$, the OTP ciphertext can be correctly decoded if the pad pairs are not excessive.

We then evaluate the YouSense's impact on collaborative sensing performance. Firstly, we suppose there are 1, 3, or 4 CR users in collaborative sensing, and for a given honest user, he collects and recovers the OTP ciphertext to obtain the combined channel decision. The channel number $M$ varies from 1 to 100, and we assume only one secure pair is adopted in $\mathbb{\overline{K}}$. We run the collaborative sensing round for 350 times, and obtain the average false positive/negative rate towards all channels. The experiment results are shown in Fig.5(a) and Fig.5(b). It is observed that when CR user senses more spectrum, the false positive/negative rate will be decreased. And in YouSense when the CR user senses more than 25 channels to recover the plaintext, the false positive/negative rate of collaborative sensing approximately equal to that in traditional collaborative sensing without YouSense. Further, we measure YouSense's influence when multiple secure pairs are adopted in $\mathbb{\overline{K}}$. The cooperator number is set as 4, and the number of secure pairs take the value of 2, 4, and 10. The experiment results are shown in Fig.5(c) and Fig.5(d). We can learn the added extra pairs requires the CR users to sense more spectrum to achieve the similar collaborative sensing accuracy. This result consists with the previous discussion that these extra pairs increase the CR user's difficulty in pad recovery, which can be adopted to thwart partial entropy selfishness.

\subsection{The Effectiveness of the YouSense}
We will demonstrate the effectiveness of YouSense when defending against entropy selfishness. We measure how much spectrum information is leaked via the ciphertext. We assume $\mathbb{\overline{K}}$ adopts a pair of pads, and the CR user randomly selects one of them to generate the OTP ciphertext. Without lose of generality, we suppose one of the pads is $\mathbf{\overline{K}}_j=[1,0,0,1]$, and measures the average report masking level of the ciphertext when the other pad $\mathbf{\overline{K}}_{j'}$ traverses all possible pads. The experiment results are given in Fig. 6. We can see when $\mathbf{\overline{K}}_{j'}$ takes $[0,1,1,0]$, the report masking level is zero. It's also clear that $\{\mathbf{\overline{K}}_j,\mathbf{\overline{K}}_{j'}\}$ is the secure pair that we've defined previously, and in this situation, the ciphertext leaks no information. This experiment result consists with our previous discussion.


\section{Related Works}

\subsection{Selfishness Issue in Traditional Ad Hoc Networks}
Selfishness in cooperation has been widely studied in traditional ad hoc networks, where the users rely on each other to forward packets through multi-hop manner. These works can be divided into two categories: reputation/credit based mechanisms. Reputation based schemes rely on individual users to monitor neighbors' traffic and keep track of each others' reputation so that uncooperative users are eventually detected and excluded from the networks\cite{reputation1,reputation2}, whereas credit-based schemes introduce virtual currency to regulate the packet forwarding relationship among different users. But these mechanisms assume the selfish behaviors could be monitored by its neighborhood. This assumption may not hold in the CRNs. In particular, an honest user cannot easily detect if a CR user is sending a fresh sensing result or not.
\begin{figure}
\centering
\includegraphics[width=0.4\textwidth]{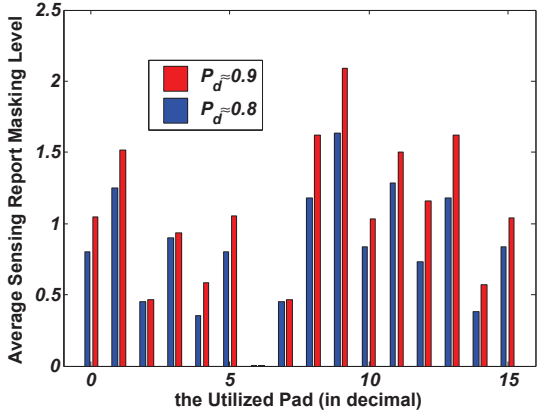}
\vspace{-0.1cm}
\caption{\textbf{The Evaluation of YouSense's Effectiveness.} In our simulation, we assume $M=4$, and the subset $\mathbb{\overline{K}}$ includes two pads. We suppose one of them $\mathbf{\overline{K}}_j$ equals to $[1,0,0,1]$, and measures the sensing report masking level when the other pad traverses all possible pads.}
\vspace{-0.5cm}
\end{figure}
\subsection{Selfishness in Collaborative Sensing}
Selfishness in collaborative sensing has received attentions recently. In \cite{qianzhang}, this problem has been firstly studied, and the incentive strategies like Grim Trigger and Carrot-and-Stick are improved to achieve better system performance. In \cite{beibei}, evolutionary game has been adopted to study how to collaborate for a CR user. Evolution Dynamics is used to analyze whether the CR user should choose to over-ride the neighbors' sensing results at the risk of no contributor and therefore no usage of the spectrum, or to contribute with some cost.
It is important to note that the current incentive schemes in collaborative sensing can work only if the selfishness can be successfully detected. However, for entropy selfishness which is difficult to identify, these schemes no longer take effects.

\section{Conclusion}
In this paper, we identify a new selfishness model named entropy selfishness in distributed collaborative sensing. Different from traditional selfish misbehavior, entropy selfishness is difficult to detect and thus can not be easily thwarted by existing incentive schemes. We further propose YouSense, a one-time pad based incentive design in which sensing reports are encrypted before sharing, to prevent the entropy selfish users from learning the sensing reports. And yet, for the honest user, he can recover this plaintext by spectrum sensing. Since the different CR users monitor the same radio spectrum set, they actually share some common observations. Then, the honest user can learn part of the plaintext by checking his own, and recover the OTP key by known plaintext attack. By this way, entropy selfishness can be mitigated in collaborative sensing. By USRP based experiment, we demonstrate that YouSense can not only prevent entropy selfishness, but also incurs little system overhead, and thus is practical to address the entropy selfishness in collaborative spectrum sensing.

\section{Acknowledgement}
This research is supported by National Natural Science Foundation of China (Grant No.61003218, 70971086, 61272444, 61161140320, 61033014, 61170267, 60934003, 60974123, 61273181 and 61221003), 973 Program (Grant No.2011CB302905), Doctoral Fund of Ministry of Education of China (Grant No.20100073120065).

\begin{appendix}\label{appendix}

\subsection{Proof of Theorem 1}
The sensing report masking level can be described as:
\begin{equation*}\label{}
    \begin{split}
    & \ \ \ \ \ \ \ I(\widetilde{C}_i, E_i^x)  \\
    & =\sum\limits_{\widetilde{C}_i\in\{0,1\}} \sum\limits_{E_i^x\in\{0,1\}}{p(\widetilde{C}_i,E_i^x) log(\frac {p(\widetilde{C}_i,E_i^x)}{p_1(\widetilde{C}_i) p_2(E_i^x)}) }
    \end{split}
\end{equation*}
In the meantime, we have:
\begin{equation}
     \begin{split}
    & p_2(E_i^x)=p(E_i^x|\widetilde{C}_i=1) p(\widetilde{C}_i=1) \\
    & +p(E_i^x|\widetilde{C}_i=-1) p(\widetilde{C}_i=-1)
     \end{split}
\end{equation}
$p(\widetilde{C}_i, E_i^x)$ is the joint probability distribution function of $\widetilde{C}_i$ and $E_i^x$. we also have:
\begin{equation}
    p(\widetilde{C}_i, E_i^x)=p(E_i^x|\widetilde{C}_i) p(\widetilde{C}_i)
\end{equation}
Suppose $\xi$ is the rate that $K_i^x$ equals to $0$, we have:
\begin{equation*}
    p(E_i^x|\widetilde{C}_i)=\xi p(R_i^x=E_i^x|\widetilde{C}_i)+(1-\xi)p(R_i^x=-E_i^x|\widetilde{C}_i)
\end{equation*}
 If $\forall \mathbf{\overline{K}}_j\in\mathbb{\overline{K}},\exists\mathbf{\overline{K}}_{j'}\in\mathbb{\overline{K}}, \mathbf{\overline{K}}_j\odot\mathbf{\overline{K}}_{j'}=\mathbf{0}$,  we could find $\xi=0.5$, then we have:
\begin{equation*}\label{}
    I(\widetilde{C}_i, E_i^x)=0
\end{equation*}

\end{appendix}

\end{document}